\documentclass[apj,twocolumn]{aastex701}

\usepackage{graphicx}
\usepackage{multirow}
\usepackage{hyperref}
\usepackage{soul}

\newcommand{\methanol}{CH$_3$OH}

\newcommand{\jb}{Jy~beam$^{-1}$}
\newcommand{\kms}{km~s$^{-1}$}

\newcommand{\jyb}{Jy~beam$^{-1}$}

\newcommand{\gtfe}{G358.93$-$0.03}
\newcommand{\gtfemmone}{G358.93$-$0.03\,MM1}

\newcommand{\vt}{$v_{t}$}

\shortauthors{Hunter et al.}

\begin{document}
\received{Nov 14, 2025}
\revised{Dec 26, 2025}
\accepted{Dec 30, 2025}

\shorttitle{Deceleration of the heat wave from the accretion outburst in G358.93$-$0.03\,MM1}
\title{A millimeter methanol maser ring tracing the deceleration of the heat wave powered by the massive protostellar accretion outburst in \gtfemmone}


\author[0000-0001-6492-0090]{T. R. Hunter}
\affiliation{National Radio Astronomy Observatory, 520 Edgemont Rd, Charlottesville, VA 22903, USA}
\email{thunter@nrao.edu}

\author[0000-0002-6558-7653]{C. L. Brogan}
\affiliation{National Radio Astronomy Observatory, 520 Edgemont Rd, Charlottesville, VA 22903, USA} 
\affiliation{Department of Astronomy, University of Virginia, P.O. Box 3818, Charlottesville, VA 22904, USA}
\email{cbrogan@nrao.edu}

\author[0000-0002-1505-2511]{G. C. MacLeod}
\affiliation{Xinjiang Astronomical Observatory, Chinese Academy of Sciences, Urumqi 830011, People's Republic of China}
\affiliation{Hartebeesthoek Radio Astronomy Observatory, PO Box 443, Krugersdorp 1740, South Africa}
\email{gord@hartrao.ac.za}

\author[0000-0001-6725-1734]{C. J. Cyganowski}
\affiliation{SUPA, School of Physics and Astronomy, University of St. Andrews, North Haugh, St. Andrews KY16 9SS, UK}
\email{cc243@st-andrews.ac.uk}

\author[0000-0003-3302-1935]{R. A. Burns}
\affiliation{RIKEN Pioneering Research Institute, 2-1 Hirosawa, Wako-shi, Saitama, 351-0198, Japan}
\email{ross.burns@riken.jp}
\affiliation{Engineering Research Institute ``Ventspils International Radio Astronomy Centre'', Ventspils University of Applied Sciences, Inženieru iela. 101, Ventspils, LV-3601, Latvia}

\author[0000-0003-1254-4817]{B. A. McGuire}
\affiliation{Department of Chemistry, Massachusetts Institute of Technology, Cambridge, MA 02139, USA}
\affiliation{National Radio Astronomy Observatory, 520 Edgemont Rd, Charlottesville, VA 22903, USA}
\email{brettmc@mit.edu}

\correspondingauthor{T. R. Hunter}
\email{thunter@nrao.edu}

\begin{abstract}
We present multi-epoch, multi-band ALMA imaging of the new Class~II millimeter methanol masers excited during the accretion outburst of the massive protostar \gtfemmone.  
The highest angular resolution image (24\,mas $\approx$ 160\,au) reveals a nearly complete, circular ring of strong maser spots in the 217.2992\,GHz (\vt=1) maser line that closely circumscribes the dust continuum emission from MM1.  Weaker maser emission lies inside the eastern and southern halves of the maser ring, generally coincident with the centimeter masers 
excited during the outburst, but avoiding the densest parts of the hot core gas traced by high excitation lines of CH$_3$CN. 
Using a variety of fitting techniques on the image cubes
of the
two strongest maser lines, each observed over 3--4 epochs, we find
the diameter of the ring increased by $\gtrsim$60\% (from $\approx$1100 to $\approx$1800\,au in the 217\,GHz line) 
over 200 days, consistent with an average radial
propagation rate of $\approx$0.01c, while the maser intensity declined exponentially.   
Fitting the angular extent of the millimeter masers versus time
yields a power law of index 0.39$\pm$0.06, which also
reproduces the observed extent of the 6.7\,GHz masers in the first VLBI epoch of \citet{Burns2020}.  This exponent is consistent with the prediction of radius vs. time in the Taylor-von Neumann-Sedov self-similar solution for an intense spherical explosion from a point source ($R\propto{t^{2/5}}$).
%
These results demonstrate the explosive nature of accretion outbursts in massive protostars and their ability to generate subluminal heat waves traceable by centimeter and millimeter masers for several months as the energy traverses the surrounding molecular material.  
\end{abstract}


\section{Introduction}

The discoveries of powerful accretion outbursts in two high-mass protostars -- NGC6334I-MM1
\citep{Hunter17,Hunter21} and S255IR-NIRS3 \citep{Caratti17,Liu18} -- provided crucial insights on the role of
episodic accretion in massive star formation  \citep{Brogan18,Cesaroni18}.  
While a study of Orion suggests that episodic accretion accounts for $\gtrsim$25\% of a star's mass \citep{Fischer19},  hydrodynamic simulations of {\bf massive} protostars predict 
values up to 40-60\% \citep[]{Meyer2021}, making it a key phenomenon to understand the process of massive star formation.
Both aforementioned outbursts were heralded by Class~II 6.7~GHz \methanol\/ maser flares in single dish observations
\citep{MacLeod18,Szymczak18,Fujisawa15},  and subsequent high resolution studies confirmed that the maser flares occurred in dense gas surrounding the outbursting protostar \citep{Hunter18,Moscadelli17}. 
Because Class~II \methanol\/ masers are pumped by infrared radiation \citep{Cragg2005}, their apparent association with protostellar luminosity outbursts has a theoretical basis.  Thus, identifying and characterizing more events is critical to understanding massive star formation.

On 2019 January 14, the 6.7~GHz Class~II \methanol\/ maser line in the massive star-forming region \gtfe\/ began flaring, rising in flux by an order of magnitude from 10~Jy \citep{Caswell10} to 99\,Jy in two weeks \citep{Sugiyama19}, and ultimately reaching 1156~Jy after seven weeks \citep{MacLeod19}.
With no prior (sub)millimeter interferometric observations, this region was known primarily as a compact clump in single dish (sub)millimeter continuum surveys, including the 1.1~mm BOLOCAM Galactic Plane Survey (BGPS) \citep[G358.936-00.032,][]{Rosolowsky10} and the 0.87\,mm ATLASGAL survey \citep[AGAL358.931-00.029,][]{Urquhart13}, with a near kinematic distance of 6.75\,kpc \citep{Brogan2019}.


%
Because the strength and rapid rise of the 6.7~GHz \methanol\/ maser flare in \gtfe\/ indicated a potential accretion outburst event, the Maser Monitoring Organization (M2O)\footnote{See the M2O website at \href{http://MaserMonitoring.org}{MaserMonitoring.org}} pursued multi-wavelength follow-up with telescopes worldwide. This effort led to the
unprecedented discovery of several never-before-seen Class\,II \methanol\/ maser lines, including some in torsionally-excited (\vt$\geq$1) states  \citep{Breen19,Volvach19,MacLeod19,Miao2022,Johnson2025}.
In March and April 2019, 14 new \methanol\/ maser lines (11 from \vt=1 and one from \vt=2) were detected by the Submillimeter Array (SMA) and the Atacama Large Millimeter/submillimeter Array (ALMA).  All the new masers showed a curvilinear spatial velocity pattern closely surrounding the dominant (sub)millimeter continuum source (MM1) in a protocluster comprised of (at least) 8 members.  VLBI observations of the well-known 6.7\,GHz maser line (\vt=0) spanning the month of February led to the discovery of a heat wave propagating outward from a location consistent with MM1, adding weight to the hypothesis of a protostellar outburst in MM1 \citep{Burns2020}.

Subsequent analysis of multi-epoch archival mid- to far-infrared data combined with new observations with SOFIA found a luminosity increase of a factor of 4.7$^{+2.1}_{-1.5}$, confirming an accretion outburst in MM1 \citep{Stecklum2021}. Of the handful of known events, the \gtfemmone\/ accretion outburst has the shortest duration \citep[see Table 3 of][]{Elbakyan2024}.  In addition,
light curves of the more common cm-wavelength \methanol\/ maser lines (\vt=0) from Hartebeesthoek Radio Observatory \citep{MacLeod19} show that the \gtfe\/ masers reached their peak in 7--10 weeks, much more rapidly than in NGC6334I-MM1 \citep[8~months,][]{MacLeod18} and S255IR-NIRS3 \citep[5~months,][]{Szymczak18}.  
In light of the results from hydrodynamic simulations of accretion in massive protostars, which predict
multiple outbursts with a range of magnitudes during the protostellar stage \citep{Meyer19a,Meyer19b}, 
it is critical to learn as much as possible from each observed event.
%
The multi-epoch cm-wavelength VLBI campaign of \citet{Burns2023} 
revealed expanding rings of methanol masers which, when combined into a single map, 
provided a sparsely sampled view of the gas kinematics within a rotating accretion disk with 
a spatial resolution sufficient to reveal spiral arms in the disk. These data also traced the propagation 
speed of the heat-wave at early times. In this paper, we show for the first time that the SMA and ALMA 
can be used to trace rings of mm-wavelength \methanol\/ maser emission which are also driven 
by the heat-wave. The size of the mm maser rings evolves with time, allowing us to measure the 
deceleration of the heat wave over 200 days.
%


\section{Observations} 


The observing parameters of our new ALMA data
are summarized in Table~\ref{maserEpochs}. 
The ALMA data were calibrated using the ALMA pipeline \citep{Hunter2023} packaged with
CASA \citep{CASATeam2022}.  In all cases, the target
was observed at the phase center, and interactive self-calibration was performed on the strongest maser line
with solutions transferred to all spectral windows with linear phase delay correction. Particularly for the longest baseline epoch (2019.451), this technique enabled unusually high-fidelity in the 25\,mas resolution 1.3\,mm continuum image.
 The SMA observations and data reduction are described in \citet{Brogan2019}. For this paper, higher
angular resolution SMA maser cubes were constructed using superuniform (SU) weighting \citep{Briggs1999}. The properties of all the image cubes are summarized in
Table~\ref{maserEpochs}.  The transition information for the two maser lines studied and 
those to which they are compared in this paper is given in Table~\ref{tab:transitions}.

\begin{deluxetable*}{cccccccccc}[hbt]
\tablecaption{Properties of the observations and the spectral line cubes\label{maserEpochs}}
\tablecolumns{10}
\setlength{\tabcolsep}{0.7mm}
\tablehead{
\colhead{Array}  & \colhead{Rest Freq.\tablenotemark{a}} & \colhead{Epoch} & \colhead{Weight} & \colhead{Beamsize} & \multicolumn{2}{c}{Peak intensity} & Peak velo. & \multicolumn{2}{c}{RMS noise: Min--Max\tablenotemark{\scriptsize d}} \\
\colhead{Config.} & \colhead{(GHz)} & (\#: year) & (robust) & \colhead{($''\times''$ (P.A. in deg))}
& \colhead{(\jb)}  & \colhead{(K)} & \colhead{(\kms)} & (mJy\,beam$^{-1}$) & (K)
} 
\startdata
\cutinhead{Details of \methanol\/ maser line cubes}
SMA vex & 199.5749 & 1: 2019.196 & SU & 0.619$\times$0.403 ($+$12) & 1203 & 149000 & $-15.52$ & 110--900 & 14--110 \\  
ALMA C4 & 199.5749 & 2: 2019.286 & SU & 0.366$\times$311 ($-$25) & 252 & 68300 & $-15.28$ & 4.7--34 & 1.3--9.1 \\ 
ALMA C8 & 199.5749 & 3: 2019.552 & 0   & 0.064$\times$0.041 (+23)\tablenotemark{\scriptsize b} &  1.1 & 12400 & $-16.60$ & 8.2--17 & 12.2--4.5 \\  
SMA vex & 217.2992 & 1: 2019.196 & SU & 0.587$\times$0.370 ($+$11) & 157  & 18700 &  $-15.52$ & 85--173 & 10--21 \\  
ALMA C10 & 217.2992 & 2: 2019.451 & 0.5 & 0.025$\times$0.021 ($-$65)\tablenotemark{\scriptsize c} & 9.0 & 450000 & $-15.88$ & 11--251 & 1.3--30 \\   
ALMA C8 & 217.2992 & 3: 2019.538 & 0.5 & 0.087$\times$0.059 ($-$25)\tablenotemark{\scriptsize c} & 8.8  & 44500 & $-15.76$ & 7.5--19 & 0.9--2.3 \\    
ALMA C6 & 217.2992 & 4: 2019.747 & $-1$  & 0.177$\times$0.124 ($-$86)\tablenotemark{\scriptsize c} & 4.8 & 5700  & $-15.76$ & 8.7--19 & 1.0--2.3 \\  
\cutinhead{Details of thermal (non-maser) line cubes}
ALMA C8 & 199.5749 & 5: 2021.599 & 0.4 & 0.076$\times$0.059 ($+$65) & 0.022 & 152 & $-17.68$ & 2.5--2.9 & 17--20\\
ALMA C8 & 217.2992 & 5: 2021.597 & 0.5 & 0.063$\times$0.052 ($+$81) & 0.018 & 146 & $-16.96$ & 1.3--1.7 & 10--13\\
ALMA C8 & 202.1648   & 5: 2021.597 & 0.4 & 0.070$\times$0.050 ($+$66) & 0.016 & 120 & $-17.68$ & 2.3--2.4& 17--18\\
ALMA C8 & 220.6793 & 5: 2021.597 & 0.5 & 0.063$\times$0.052 ($+$80) & 0.015 & 115 & $-17.32$ & 1.7--1.9 & 13--15\\
ALMA C8 & 220.5393 & 5: 2021.597 & 0.5 & 0.063$\times$0.052 ($+$80) & 0.016 & 121 & $-17.68$ & 1.6--1.8 & 12--14\\
\enddata
\tablenotetext{a}{Final three rows are CH$_3$CN: $J$=11--10 $K$=7, $J$=12--11 $K$=4,   and $J$=12--11 $K$=7, respectively.}
\tablenotetext{b}{Prior to the 2-component fitting process, this cube was smoothed to the beamsize of the 2019.286 epoch (see \S~\ref{fitResults}).}
\tablenotetext{c}{Same as $b$ but smoothed to the beamsize of the SMA epoch (see \S~\ref{fitResults}).}
\tablenotetext{d}{Range of RMS per channel where channel width = 0.12\,\kms\/ (ALMA \methanol), 0.24\,\kms\/ (SMA), and 0.36\,\kms (CH$_3$CN).}
\end{deluxetable*}


\begin{deluxetable}{rcrc}[hbt]
\tablecaption{Properties of the \methanol\/ maser transitions \label{tab:transitions}}
\tablecolumns{4}
\setlength{\tabcolsep}{1.3mm}
\tablehead{
\colhead{Frequency\tablenotemark{a}}  & \colhead{Symmetry \& Transition} & \colhead{$E_{\rm up}/k$} & \colhead{Ref.\tablenotemark{b}}\\
\colhead{(GHz)}  & \colhead{$\sigma, J(K)$ -- $J'(K')$, \vt} & \colhead{(K)} & 
}%
\startdata
\cutinhead{observed lines}
199.574851(18) & $E_2$, 13(-2) -- 14(-3), \vt=1 & 575.18 & 1\\
217.299205(17) & $A^-$, 6(1) -- 7(2), \vt=1 & 373.92 & 1 \\
\cutinhead{compared lines}
6.6685192(8) & $A^+$, 5(1) -- 6(0), \vt=0 & 49.06 & 2\\
20.970620(21) & $A^+$, 10(1) -- 11(2), \vt=1 & 452.44 & 1
\enddata
\tablenotetext{a}{Uncertainty in final digit(s) given in parentheses.}
\tablenotetext{b}{References: (1) CDMS \citep{CDMS2016}, (2) \citep{Muller2004} }
\end{deluxetable}

\section{Results}

\begin{figure*}[th]   %
\centering
\newcommand{\mywidth}{0.46} 
\includegraphics[width=\mywidth\linewidth]{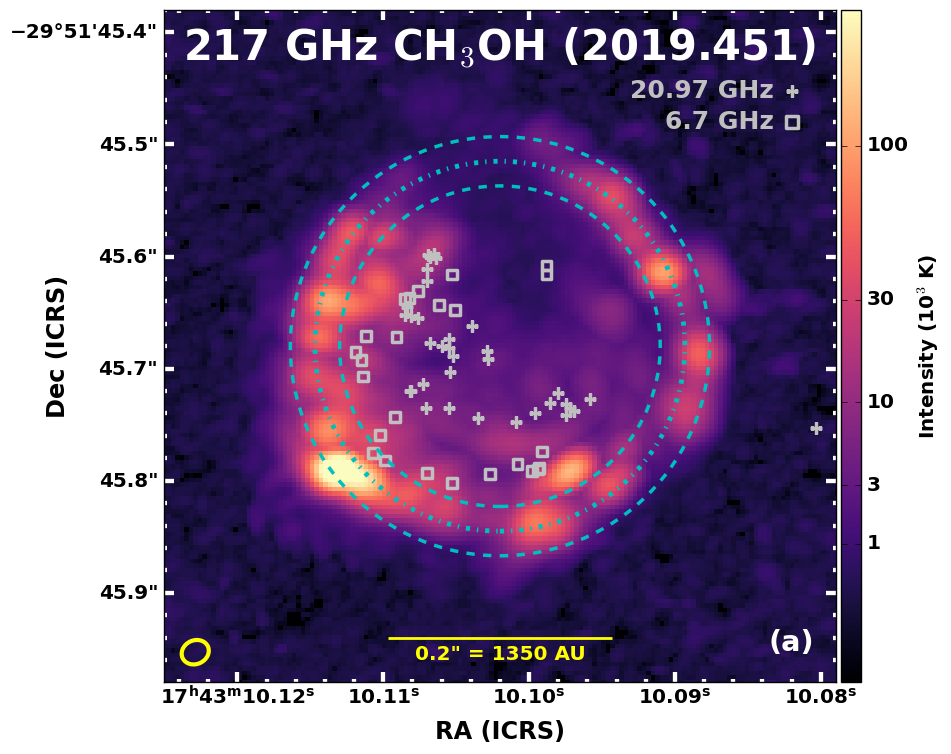} 
\includegraphics[width=\mywidth\linewidth]{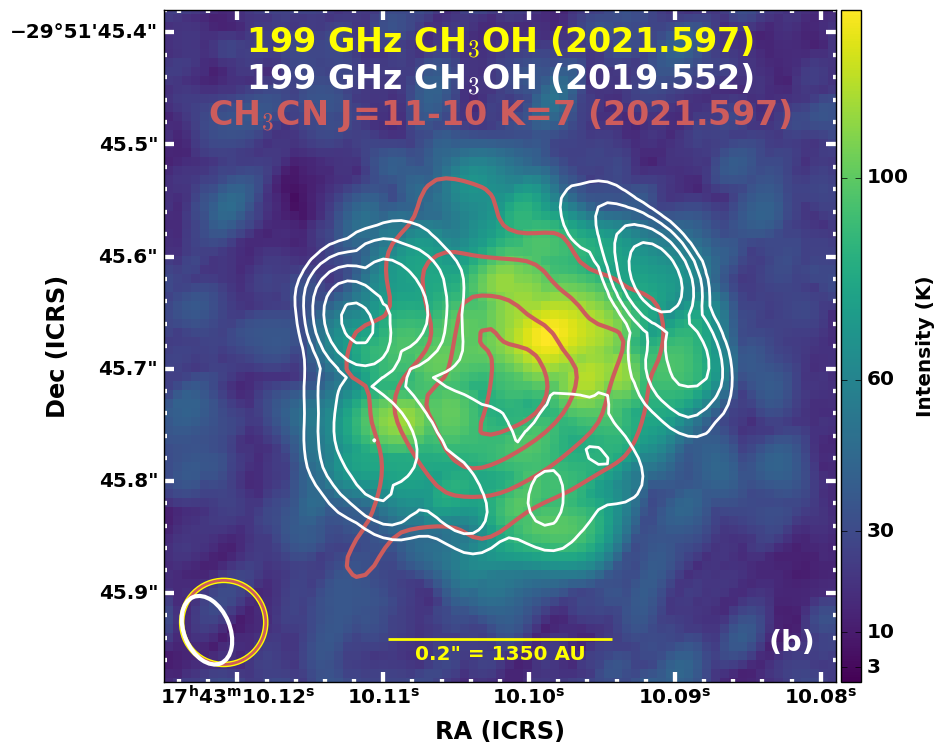} 
\includegraphics[width=\mywidth\linewidth]{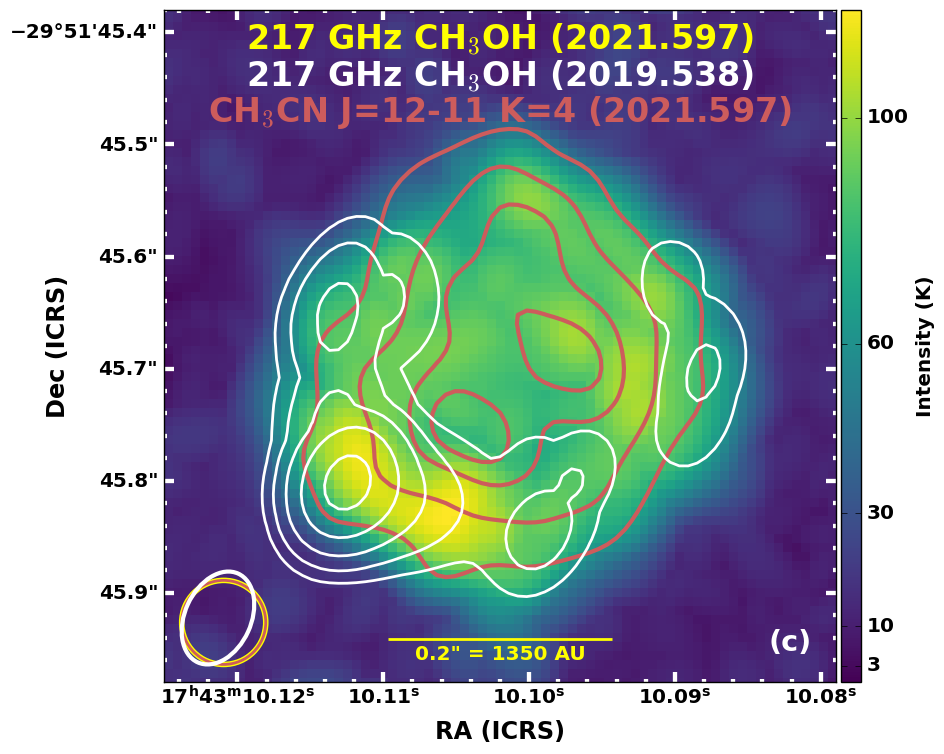} 
\includegraphics[width=\mywidth\linewidth]{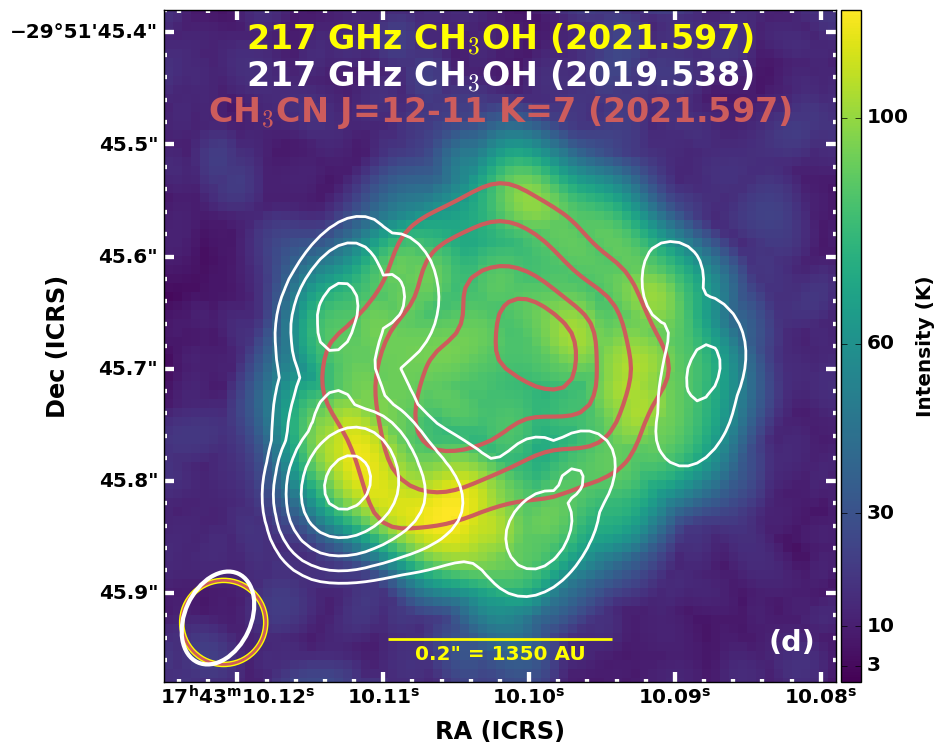} 
\includegraphics[width=\mywidth\linewidth]{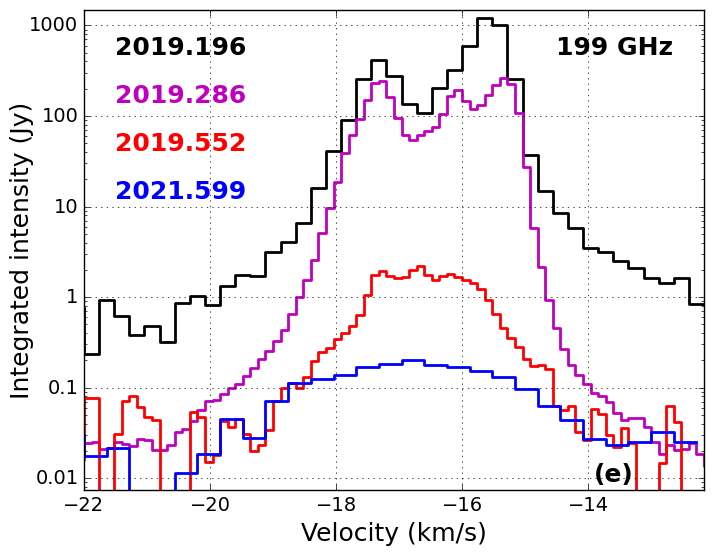} 
\includegraphics[width=\mywidth\linewidth]{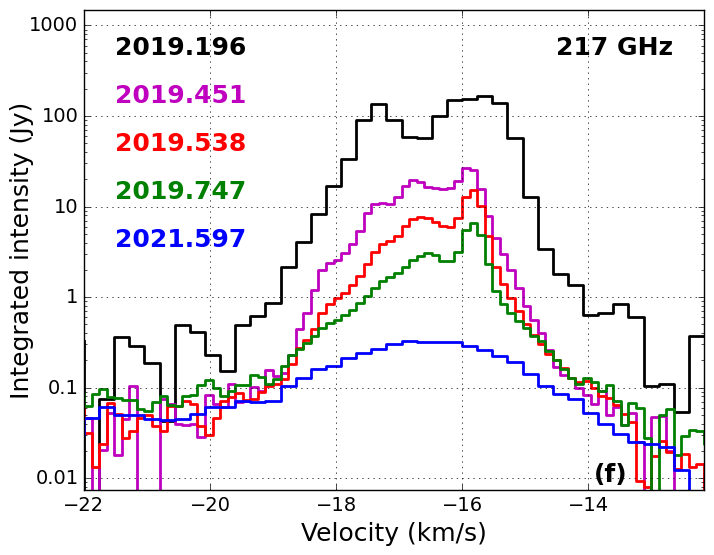} 
\caption{(a) Peak intensity image of the 217\,GHz masers surrounding MM1 at Epoch 2019.451 with squares marking 6.7\,GHz maser spots at Epoch 2019.429 from \citet{Burns2023} and $+$ signs marking 20.9706\,GHz maser spots at Epoch 2019.421 from \citet{Bayandina2022}. The blue dash-dotted circle is a fit to the Gaussian-fitted positions of the 8 strongest spatial components around the ring, while the dashed circles represent the 1$\sigma$ uncertainty on the fit. Panels (b-d) show in colorscale the peak intensity of the 199 GHz (b) and 217\,GHz  (c, d) \methanol\/ transitions, smoothed to $0\farcs075$ resolution, from the 2021.6 epoch data when the emission is consistent with thermal excitation. White contours show the peak intensity of the 199 GHz (b) and 217 GHz (c, d) \methanol\/ lines while they were still strongly masing circa 2019.5 (levels: 5, 10, 20, 40, 80\% of the maser peak). Also overlaid on (b-d) are red contours of the CH$_3$CN integrated intensity, smoothed to $0\farcs075$ resolution, for the J=11-10 K=7, J=12-11 K=4, and J=12-11 K=7 transitions, respectively (levels: 25, 50, 75, 90\% of the peak). Panels (e-f) show spectra of the central 10\,\kms\/ spatially integrated over the region of significant emission at each of the epochs from Table~\ref{maserEpochs}.
\label{figure1}}
\end{figure*}

\subsection{Maser morphology at highest resolution}
The sequence of ALMA observations were obtained over many months as the antennas were moved to larger then smaller configurations.
The highest angular resolution image of the millimeter masers in \gtfe\/ (160\,au) was obtained in epoch 2019.451, in the 217.2992\,GHz $A$-type \methanol\/ line. 
The 
peak intensity (moment 8) image reveals a striking ring-like structure that circumscribes the dust continuum emission, which has an extent of 0$\farcs$24 $\sim$1600\,au (Fig.~\ref{figure1}a, \ref{maserMontage}b).   The ring is nearly complete except on the northern edge, where the gap contains a secondary peak of continuum emission.
Two cm maser lines of $A$-type \methanol\/ were observed within a few days of this image, 6.668\,GHz (\vt=0) 
with VLBI \citep{Burns2023} and 20.97062\,GHz (\vt=1) 
with the Karl G. Jansky Very Large Array \citep{Bayandina2022}, and those spots all lie
inside the ring of strongest 217\,GHz maser emission.  Extended 217\,GHz maser emission is present inside the ring (Fig.~\ref{figure1}a) with a brightness temperature still well above the upper level excitation energy of this transition 
($E_{\rm up}$=374\,K).
In many channels, 217\,GHz maser emission appears on both sides of the ring, which will produce blending in lower resolution images. In hindsight, fitting a single spatial component to such images  \citep[as in][]{Brogan2019} will lead to biased estimates of the emission extent.  Therefore, the extent of the fitted pattern of masers in the low resolution epochs must be revisited (see \S~\ref{technique}).

\subsection{Exponential decay into thermal emission}

The strongest \methanol\/ maser lines observed in $\geq3$ epochs were the 199 and 217\,GHz lines, thus we use their data in our analysis. 
In both cases, the first epoch is from SMA, which had the coarsest angular resolution, while the rest are from ALMA (Table~\ref{maserEpochs}). The first epoch corresponds to 14\,days after the second 6.7\,GHz epoch of \citet{Burns2020}, whereupon the 6.7\,GHz maser extent had reached a diameter of $\sim$1000\,au \citep{Burns2020}.  It is also only 2~days after the 12.178 and 6.7\,GHz masers had reached their peak single-dish intensities \citep{MacLeod19}. Both millimeter maser lines (199 and 217\,GHz) decayed exponentially (Fig.~\ref{expansion}a), with fitted $e$-folding times of 19.7$\pm$0.38 and 61.7$\pm$7.4 days, respectively, for their peak intensities, and 20.0$\pm$0.5 and 44.9$\pm$8.5~days for their integrated flux densities.  The 217\,GHz line shows a significantly slower rate of decline, making it detectable for the longest duration among the five millimeter methanol maser lines observed by SMA or ALMA for at least 3 epochs during 2019\footnote{The other three lines are the 200, 201, and 215 GHz \vt=1 lines; see Table 3 of \citet{Brogan2019} for their frequencies and $E_{\rm up}$ values.}.  Exponential decay of \methanol\/ maser emission was also observed during the S255IR-NIRS3 outburst \citep[6.7 GHz,][]{Szymczak18} and in single-dish observations of \gtfe:  six lines in the 3\,mm band \citep{Zhang2024} and two lines of $^{13}$\methanol\/ in Ku band \citep{Chen2020}.

By the time of our ALMA observations on 2021.597, the 199 and 217\,GHz lines
were considerably weakened, with peak intensities 
similar to the other bright thermal hot core lines in MM1 (Table~\ref{maserEpochs}).
Their brightness temperatures were significantly below the excitation energy levels of these transitions (Table~\ref{tab:transitions}), and their line profiles were Gaussian shaped in the central 5\,\kms\/ instead of the prior multi-peaked profile (Figure~\ref{maserEpochs}e,f).  These two attributes indicate
that these transitions were no longer masing.
The morphology of the residual thermal \methanol\/ emission (Fig.~\ref{figure1}b,c,d) spans essentially the same diameter as the maser emission observed two years prior, though the spatial peaks generally
do not coincide. Other hot core species tracing dense gas with a similar $E_{\rm up}$ as the 217\,GHz line, such as CH$_3$CN $K$=7 from $J$=11--10 and $J$=12-11 (Fig.~\ref{maserMontage}b,d), arise mostly inside the (former) maser ring.   The strongest emission of the higher-lying 199\,GHz thermal line ($E_{\rm up}$=575\,K) originates from close to the $K$=7 peaks ($E_{\rm up}$=408--419\,K).
While the extents of these CH$_3$CN lines are comparable to thermal \methanol, their peak intensities are not correlated with the coeval thermal or (prior) masing \methanol\/ gas. 

\subsection{Maser position fitting technique}
\label{technique}

\begin{figure*}[th]   %
%
%
\centering
\newcommand{\mywidth}{0.49} 
%
\includegraphics[width=\mywidth\linewidth]{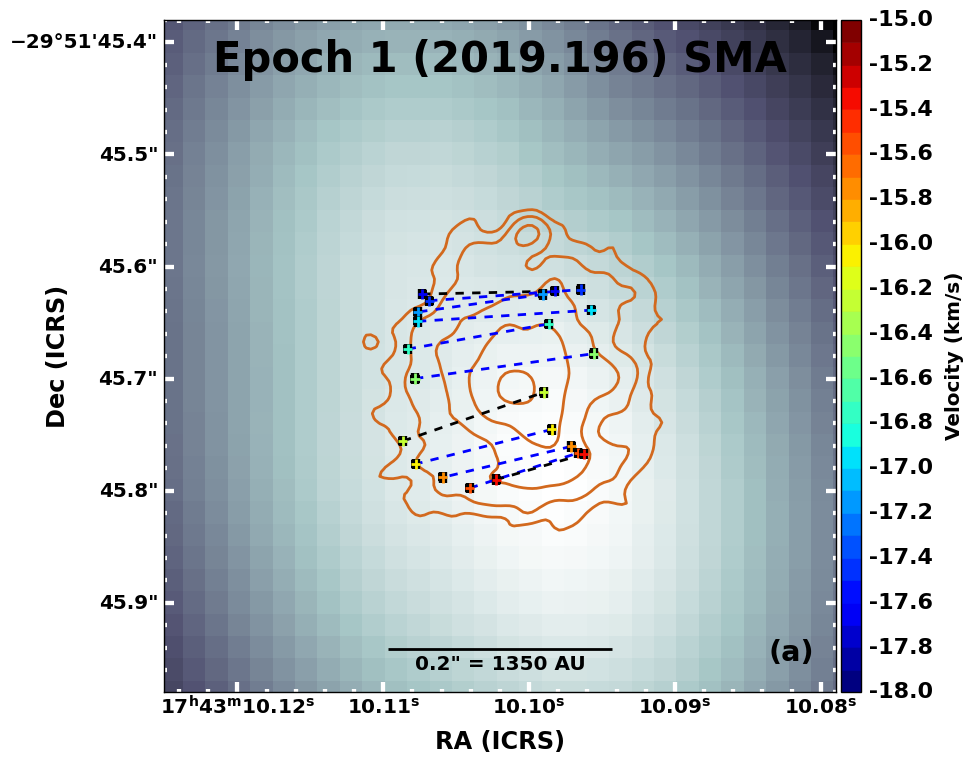}
%
\includegraphics[width=\mywidth\linewidth]{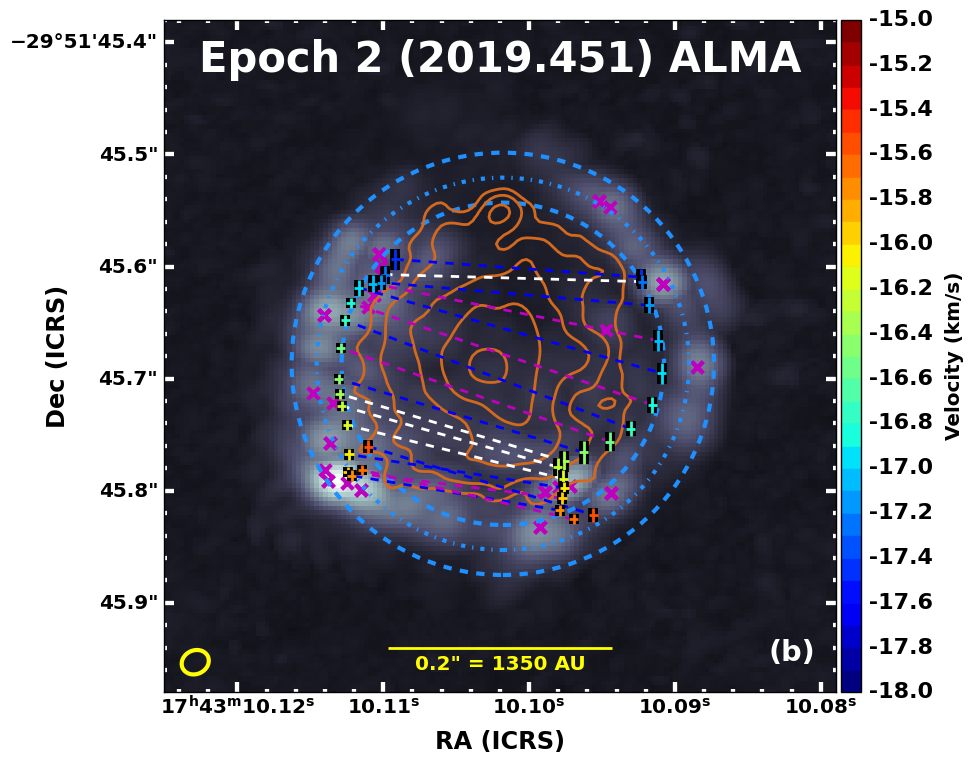}
%
%
%
%
%
\includegraphics[width=\mywidth\linewidth]{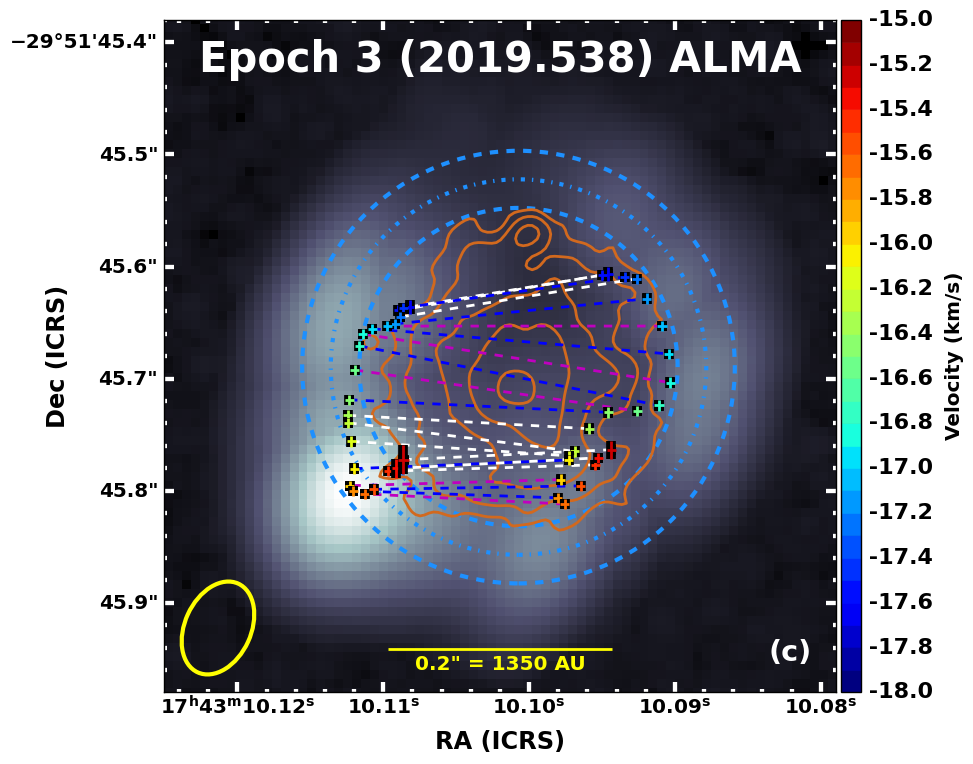}
%
\includegraphics[width=\mywidth\linewidth]{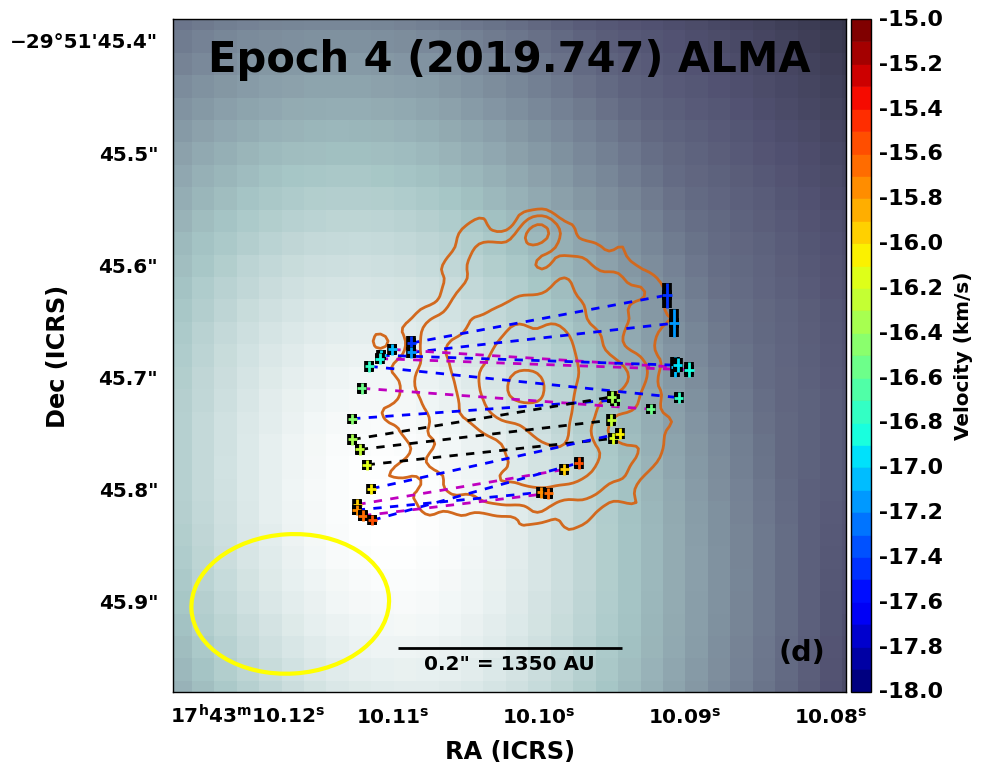} 
\caption{Native resolution peak intensity images of the 217\,GHz maser emission in greyscale, ordered by epoch, overlaid with 217\,GHz continuum contours from epoch 2019.538 (0.05, 0.1, 0.2, 0.4, 0.8~\jyb), and marked with the 2-component fits as crosses colored by velocity and connected to their counterparts by dashed lines. The initial estimates are shown as magenta X's in panel b. Blue dashed lines denote channels fit successfully across all epochs, magenta dashed lines denote channels fit successfully across only the ALMA epochs, while white and black dashed lines denote the remaining channels fit successfully in two or fewer epochs. The blue circles in panels b and c are the circle fits (\S~\ref{technique}) for their epochs, with the dashdot circle being the best fit and the dashed circles representing the 1$\sigma$ uncertainties.
}
\label{maserMontage}
\end{figure*}

Image cubes from three ALMA epochs have sufficient angular resolution to independently fit several maser spots around the ring.  To measure the ring diameter, we subsequently fit those positions with a circle using {\tt scipy.odr} accounting for the uncertainties of those fitted positions, with results summarized in Table~\ref{tab:fitResults} and Fig.~\ref{maserMontage}b,c.  The offsets of the centers of the fitted circles
from the continuum peak range from 7--16\,mas (50-110\,au).
In the other four epochs, although the ring is not clearly resolved, the emission is extended east-west.  Fortunately, even the lowest angular resolution SMA cubes from the first epoch, when the maser lines were strongest, allow two spatial
components to be fit reliably in channels with sufficiently high S/N and thereby provide
a potential method to measure changes in the structure.  

To establish the effectiveness of 2-component
fitting of this structure, we performed a proof-of-concept test by smoothing the highest resolution ALMA
cube of the 217.2992\,GHz line (i.e., the second epoch: 2019.451) to the much lower resolution of the 
SMA epoch (Table~\ref{maserEpochs}). First, we examined the native resolution ALMA cube and marked two spots in each channel: the brightest spot in each of the eastern and western halves of the ring.  
We ran the CASA task {\tt
imfit} to fit two Gaussians to each channel of the smoothed cube 
using these initial position estimates.
We fixed the size and position angle of each component to the synthesized beamsize and solved for their positions and intensities. In the channels where the fitting task succeeded mathematically, we examined the locations of the fitted components and the residual images and established several criteria to determine if a 2-component fit was robust and reliable. First, the peak absolute intensity in the residual image of the 2-component fit must be $<0.4$ times that of a single-component fit to the same channel.  Second, the integrated absolute flux in the residual image of the 2-component fit must be $<0.6$ that of a single-component fit.  These two requirements demonstrate that a 2-component fit is significantly more effective in modeling the total flux.  Third, the separation of the fitted components in right ascension must be between $0\farcs09$ and $0\farcs40$.  The lower limit avoids solutions where the two components are nearly coincident (i.e. on the same side of the ring) and the upper limit avoids anomalous fits far outside the ring. 
To check the fit robustness, 2-component fits were also run with the initial estimates set to two locations separated by 0\farcs18 straddling
the single component fit; we confirmed that nearly identical fit results were still obtained. 
To check the fit reliability, we overlaid the reliable 2-component fit results on the peak intensity image of the native-resolution parent data, demonstrating that the fitted positions generally follow the true underlying structure in the two highest resolution ALMA epochs (Fig.~\ref{maserMontage}b,c) to within the native beamsize.  However, the 2-component fit
results systematically tend to agree with the $-1\sigma$ diameter from the ring fit rather than
the fitted diameter, hence the results from these two distinct methods should not be combined when fitting for
the time variation.

Assuming the ring-like structure was pervasive in time, we performed the same spatial smoothing 
to the SMA cube resolution for the other ALMA epochs at 217.2992\,GHz, followed by 2-component fitting using the same initial estimates as epoch 2019.451 with the same process and success criteria. This method ensures a globally self-consistent fitting approach across epochs and observatories while not imposing any assumption of a smaller diameter at earlier epochs. The same method was applied to the 199.5749\,GHz line, but the maser was too weak in the final ALMA epoch to achieve 2-component fits after smoothing it to the SMA resolution.  Therefore, we smoothed the final ALMA epoch cube to the lower resolution ALMA epoch and fit that cube with the same process.   Finally, we fit the SMA 
cubes for both lines with this process, applying the astrometric shifts required \citep{Brogan2019}.

\subsection{Maser fitting results}
\label{fitResults}

The successful 2-component fit results from both observatories for the 217\,GHz line are illustrated in Figure~\ref{maserMontage}.
The lengths of the lines connecting the two components in each channel
reveal a trend of increasing separation with time.
Within each epoch, a range of separations is found, in part
due to the velocity structure of the ring where different chord lengths are traced in different channels as the emission moves along the ring (mostly in declination).  Since the diameter is the largest chord of a circle, we take the 90th percentile measurement as the best estimate of the diameter, with the estimated uncertainty being the difference between it and the 80th and 100th percentile values. To reflect the additional uncertainty of fitting relative separations smaller than the
observed beamsize ($\theta$), we add in quadrature the ratio of $\theta$ to the median SNR of the individual separation measurements.

\begin{figure*}[th!]   %
\centering
\includegraphics[width=0.62\linewidth]{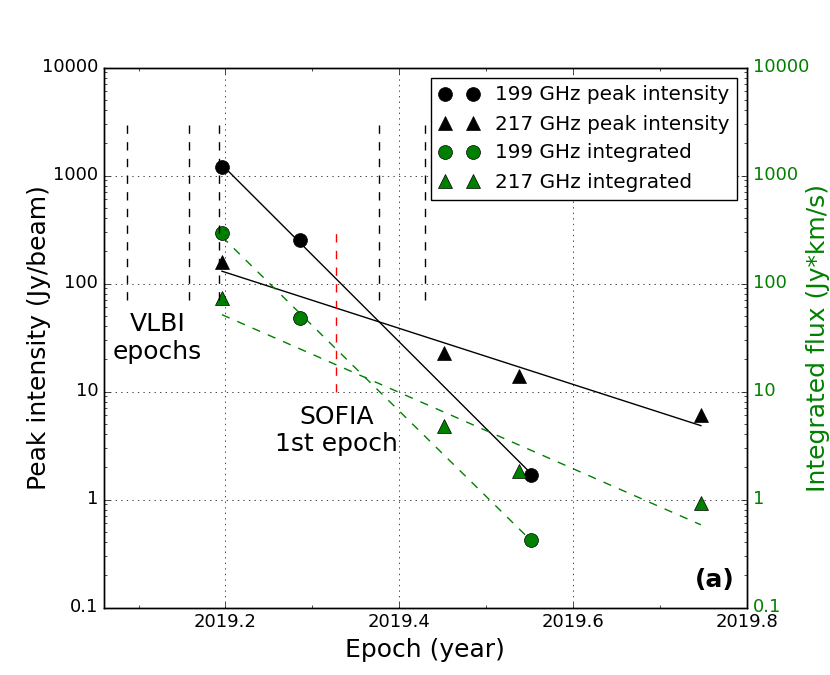}   
\includegraphics[width=0.62\linewidth]{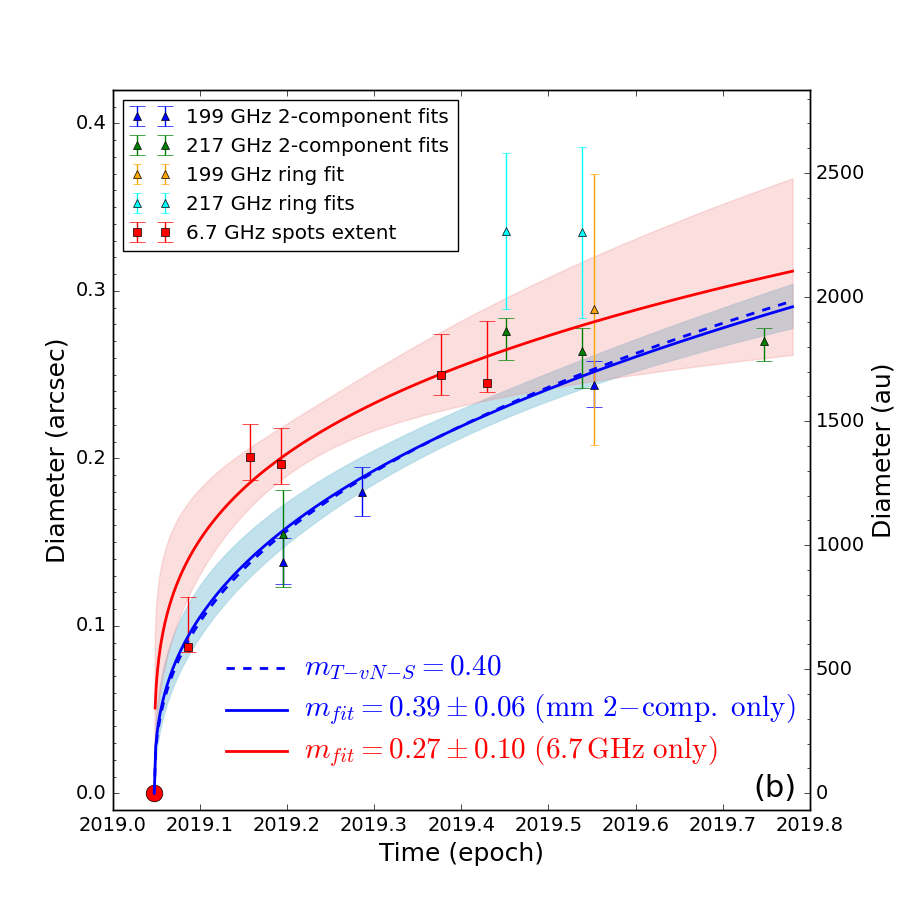}  
\caption{(a) Light curve of the 199 and 217\,GHz maser lines on a semi-log scale overlaid by fitted models of exponential decay, with vertical dashed lines denoting the VLBI epochs \citep{Burns2020} and first SOFIA epoch \citep{Stecklum2021} for reference;  (b) Ring diameter vs. time as inferred from the various
maser tracers and methods described in \S~\ref{technique}.  The solid curves are fractional power-law fits to the millimeter maser 2-component fit results from this work (blue) and independently to the 6.7\,GHz VLBI measurements (red) from \citet{Burns2023}. The best-fit blue solid curve is consistent with the Taylor-von Neumann-Sedov exponent (blue dashed curve) within the 
90\% confidence region indicated by the blue shaded area computed from bootstrap analysis (see Section~\ref{fitResults}). The red 
spot marks the inception of the maser flare \citep{Sugiyama19}. The gap in the red curve near the origin is where the fit would exceed the speed of light.
}
\label{expansion}
\end{figure*}

The fit results for each epoch of each line are
listed in Table~\ref{tab:fitResults} and illustrated in Figure~\ref{expansion}b.  The results indicate a general trend that the emission diameter increased with time.
Also shown in Fig.~\ref{expansion}b are measurements of
the diameter of the distribution of 6.7\,GHz maser spots fit to
5 epochs of VLBI images by \citet{Burns2023}.  Here, the diameter is taken as the smallest circle containing 90\% of the fitted spots and the uncertainties correspond to the smallest circles containing 80\% and 100\% of the spots.
Fitting the inferred diameters from the 2-component fits to the millimeter maser images using a fractional power law, $R=a(t-t_0)^m$, assuming $R$=0 at $t$=$t_0$=2019.0479 corresponding to the maser flare inception \citep{Sugiyama19} yields $m$ = 0.39$\pm$0.06.  The width of the 90\% confidence region of the fit shown in Fig.~\ref{expansion}b corresponds to
the interdecile range of $50000$ Monte Carlo trial fits with bootstrap resampling.  This fit is consistent with the size measurement of the first epoch of 6.7\,GHz masers.  However, an independent fit to all five epochs of the 6.7\,GHz maser spot extent measurements, including four epochs obtained amidst the millimeter maser measurements, produces a shallower power law index. These results demonstrate that one should be cautious and not simply combine size measurements from
maser transitions that likely exhibit population inversion in different ranges of physical conditions, particularly when detailed theoretical models are not available for all transitions.  In this case, the 6.7 GHz line commonly observed as a strong maser in many high-mass star formation regions \citep{Caswell10} is known to show population inversion for a wide range of physical conditions \citep{Cragg2005}. In contrast, the rarer transitions like the 199 and 217\,GHz lines likely show significant population inversion for a relatively narrower range of physical conditions, only (to date) observed in accretion bursts. 

\begin{deluxetable}{ccccc}[ht]
\tablecolumns{5}
\setlength{\tabcolsep}{1.4mm}
\tablecaption{{Summary of maser fitting results}\label{tab:fitResults}}
\tablehead{
\colhead{Epoch} & \colhead{Frequency} & No. spots & \multicolumn{2}{c}{Inferred diameter} \\
\colhead{(year)} & \colhead{(GHz)} & & (arcsec) & (AU)
}
\startdata
\cutinhead{ring fitting of high resolution ALMA cubes}
2019.451 & 217.2992 & 8 & 0.336$\pm$0.047 & 2270$\pm$310\\ 
2019.538 & 217.2992 & 6 & 0.335$\pm$0.051 & 2265$\pm$345\\ 
2019.552 & 199.5748 & 7 & 0.289$\pm$0.081 & 1950$\pm$550\\ 
\cutinhead{2-component fitting of SMA and ALMA cubes}
2019.196 & 199.5749 & 7 & $0.138^{+0.014}_{-0.013}$ & $930^{+100}_{-90}$\\
2019.196 & 217.2992 & 11 & $0.155^{+0.026}_{-0.032}$ & $1050^{+180}_{-220}$\\
2019.286 & 199.5749 & 21 & $0.180^{+0.015}_{-0.014}$ & $1210^{+100}_{-100}$\\
2019.451 & 217.2992 & 17 & $0.276^{+0.008}_{-0.017}$ & $1860^{+50}_{-120}$\\
2019.538 & 217.2992 & 22 & $0.264^{+0.014}_{-0.022}$ & $1780^{+100}_{-150}$\\
2019.552 & 199.5749 & 10 & $0.244^{+0.014}_{-0.013}$ & $1650^{+100}_{-90}$\\
2019.747 & 217.2992 & 16 & $0.270^{+0.008}_{-0.012}$ & $1820^{+60}_{-80}$\\
\cutinhead{ring inferred from extent of VLBI spot data}
2019.086 & 6.668 & 140 & $0.080^{+0.030}_{-0.003}$ & $540^{+205}_{-20}$ \\
2019.158 & 6.668 & 219 & $0.187^{+0.020}_{-0.013}$ & $1260^{+133}_{-91}$ \\
2019.193 & 6.668 & 192 & $0.192^{+0.021}_{-0.012}$ & $1297^{+143}_{-81}$ \\
2019.377 & 6.668 & 70 & $0.237^{+0.025}_{-0.012}$ & $1602^{+166}_{-83}$ \\
2019.429 & 6.668 & 75 & $0.235^{+0.037}_{-0.005}$ & $1584^{+252}_{-36}$ \\
    \enddata
\end{deluxetable}

\section{Discussion}

\subsection{Propagation of the heat wave}

In the regime of low dust opacity, the heat wave from an accretion outburst propagates outward essentially at the speed of light as thermal energy carried by photons is promptly absorbed by dust grains and re-radiated at longer wavelengths \citep{Johnstone13}. However, at high dust optical depths, the photon mean free path drops and radiative transfer
is slowed \citep{WolfPhDT}.  Evidence for this effect around massive protostars was seen first in the outburst of S255IR-NIRS3 where the delay between the rise of pre-existing and new maser components \citep{Moscadelli17} implied a propagation speed of $\sim$0.15c \citep{Stecklum2018}.
The increase in size of the maser ring around \gtfemmone\/  (Fig.~\ref{expansion}b) follows a similar trend to the 6.7\,GHz VLBI observations, which show an apparent propagation rate of $\sim$0.08c during February 2019 when the diameter increased from $\sim$540\,au to $\sim$1260\,au \citep{Burns2020}. 
We interpret our new results as tracing the subsequent outward progress of the heat wave over the 14--215 day interval following those February epochs.  
 Based on the millimeter maser data, the extent of the maser emission increased by $\gtrsim$60\% ($\sim$1100\,au to $\sim$1800\,au in the 217\,GHz line), corresponding to a mean radial expansion velocity of 0.01$c$.  
The power law fit to the 2-component fits to the millimeter maser cubes is consistent with the prediction of radius vs. time by the Taylor-von Neumann-Sedov self-similar solution for an intense spherical explosion from a point source in a uniform density medium \citep[$R\propto{t^{2/5}}$,][]{Taylor1955,Sedov1946}.  The protostar MM1 is the obvious source of radiative energy as its location, inferred from the peak dust emission, is at the center of the maser ring (Fig.~\ref{maserMontage}b).   

The good agreement with a model based on uniform gas density is not unreasonable, because the torsionally-excited millimeter \methanol\/ masers, which are tracing the propagation, are likely excited only within a particular range of density and column density. . The dearth of masers toward the peak of the warmest thermal gas tracer (199\,GHz, Fig.~\ref{figure1}b) likely indicates the density there is high enough to collisionally quench them.
Furthermore, the limited extent of thermal emission in epoch 2021.597 (Fig.~\ref{figure1}b,c,d) suggests that
the maser flare subsidence corresponds to when the heat wave
reached a radius where the gas density was too low to sufficiently excite \methanol.
Regardless, these results provide direct evidence that the heat wave propagated for at least 9 months, decelerating as the explosive nature of the energy released by the accretion outburst traversed the surrounding molecular material.

\subsection{Connection to the infrared continuum decay and the accretion event}
\label{conclusion}

The high line intensities seen during the outburst 
from \gtfemmone\/ indicate these masers are saturated.  In this regime, their
intensity scales linearly with the maser optical depth, which in turn scales linearly with the
pumping radiation \citep{Moscadelli17}.  Class~II \methanol\/ masers are pumped by mid-infrared
radiation \citep{Cragg2005}, which should lead to a close coupling between the maser intensity and the 
combined luminosity of the protostar and its accretion disk.
Evidence for this coupling has been found in the intermediate mass young stellar object (YSO) G107.298+5.639 \citep{Olech2020}. 
Examples of exponential decay of the infrared continuum have been seen in a variety of outbursting YSOs, from
massive \citep[G323.46$-$0.08,][]{Wolf2024} to low-mass
\citep[e.g.,][]{Wang2023},
and in models of the dust emission from 
expanding supernova remnants \citep{Drozdov2025}.  
In the context of protostellar accretion, magnetohydrodynamic 
simulations of low-mass protostars with high magnetic fields (2\,kG) produce accretion outbursts with exponentially decaying light curves \citep{Gaches2024}.  This behavior is due to the magnetic Rayleigh-Taylor 
fluid instability developing at the radius where the magnetic field truncates the disk leading to
magnetospheric accretion \citep{Zhu2024}.  

Building on these ideas, one interpretation of the \gtfemmone\/ event is that the mass ultimately accreted \citep[modeled to be 180\,$M_\earth$,][]{Stecklum2021} entered the existing disk resulting in disk instability and leading to
a spike in the accretion rate which initiated the heat wave. The relatively short duration
of this outburst argues 
against bloating of the protostellar photosphere as a significant contributor to the luminosity increase. 
Because it remains unclear whether the disk or the protostar dominates the total luminosity in massive YSO outbursts, we cannot say whether the subsequent accretion rate remained steady or declined, only that the total luminosity of the disk plus protostar did eventually decay.  
Similar high resolution observations and analysis of future events, 
especially with higher cadence monitoring in the infrared and (sub)millimeter, will help to
understand the prevalence, behavior, and  underlying mechanisms driving these heat waves.
%
%
Finally, the relatively long exponential decay timescales of the 217\,GHz and 93.196~GHz maser lines (61.7$\pm$7.4 and 66.6$\pm$11.1 days, respectively; this work and \citet{Zhang2024}), suggests these lines may offer the most promise of detection during outbursts in other massive star forming regions.  Interestingly, these maser lines lie within 200\,MHz of SiO (5--4) and within 25\,MHz of N$_2$H$^+$ (1--0), respectively, so there should be a substantial amount of data in the ALMA archive that could be searched for their fleeting appearances in other objects.

\begin{acknowledgments}

The National Radio Astronomy Observatory is a facility of the National Science Foundation operated under agreement by the Associated Universities, Inc. This paper makes use of the following ALMA data: ADS/JAO.ALMA\#2018.A.00031.T and 2019.1.00768.S. 
ALMA is a partnership of ESO (representing its member states), NSF (USA) and NINS (Japan), together with NRC (Canada) and NSC and ASIAA (Taiwan) and KASI (Republic of Korea), in cooperation with the Republic of Chile. The Joint ALMA Observatory is operated by ESO, AUI/NRAO and NAOJ.    
This research made use of NASA's Astrophysics Data System, funded by NASA under Cooperative Agreement 80NSSC21M0056.
The SMA is a joint project between the Smithsonian Astrophysical Observatory and the Academia Sinica Institute of Astronomy and Astrophysics and is funded by the Smithsonian Institution and the Academia Sinica.
This research used the \url{https://www.splatalogue.net} spectroscopy database \citep{Remijan07}.
C.J.C. acknowledges support from the STFC (grant ST/Y002229/1).
R.A.B acknowledges support from the Latvian Council of Science project ``A single-baseline radio interferometer in a new age of transient astrophysics (IVARS)'' (Grant No. lzp-2022/1-0083).
The authors wish to recognize and acknowledge the very significant cultural role and reverence that the summit of Maunakea has always had within the indigenous Hawaiian community.  We are most fortunate to have the opportunity to conduct observations from this mountain.  We thank the anonymous referee for useful suggestions and recommendations that improved the results and analysis in this paper.

\end{acknowledgments}

\facility{ALMA, SMA}
\software{CASA \citep{CASATeam2022}, ALMA pipeline \citep{Hunter2023}, Astropy \citep{astropy:2013, astropy:2018, astropy:2022}, APLpy \citep{aplpy2012}}, NumPy \citep{numpy2020}, SciPy \citep{scipy}

\bibliography{bibliography}
\bibliographystyle{aasjournalv7}

\end{document}